\documentclass[aps,preprint, superscriptaddress]{revtex4-1}
\usepackage{amsmath}
\usepackage{epstopdf}
\usepackage{graphicx}
\usepackage{color}
\usepackage{float}

\begin{document}

\pdfoutput=1

\title{Suppression of charge density wave phases in ultrathin 1T-TaS$_2$}

\author{Adina Luican-Mayer}
\affiliation{Center for Nanoscale Materials, Argonne National Laboratory, Argonne, Illinois 60439, United States}

\author{Jeffrey R. Guest}
\affiliation{Center for Nanoscale Materials, Argonne National Laboratory, Argonne, Illinois 60439, United States}

\author{Saw Wai Hla}
\affiliation{Center for Nanoscale Materials, Argonne National Laboratory, Argonne, Illinois 60439, United States}

\date{\today}

\begin{abstract}

Using temperature dependent Raman spectroscopy we address the question of how the transition from bulk to few atomic layers affects the charge density wave (CDW) phases in 1T-TaS$_2$. We find that for crystals with thickness larger than $\approx 10nm$ the transition temperatures between the different phases as well as the hysteresis that occurs in the thermal cycle correspond to the ones expected for a bulk sample. However, when the crystals become thinner than $\approx 10nm$, the commensurate CDW phase is suppressed down to the experimentally accessible temperatures. In addition, the nearly commensurate CDW phase is diminished below $\approx 4nm$. These findings suggest that the interlayer coupling plays a significant role in determining the properties of  CDW systems consisting of a few unit cells in the vertical direction. 
\end{abstract}

\maketitle

Transition metal dichalcogenides are typically layered materials, TX$_2$ (T= transition metal, X=chalcogen: S, Se,Te) built out of stacked three-layers units (X-T-X) coupled by weak van der Waals forces. Their bulk electrical properties span a wide range: from insulating (HfS$_2$), semiconducting (MoS$_2$) to metallic showing superconductivity or charge density order (e.g. NbS$_2$, TaS$_2$, TiSe$_2$). The competition of various effects such as dimensionality, electron-phonon coupling, electron-electron interactions, and disorder gives rise to the wealth of phases in these materials \cite{wilyof}. The anisotropy of the crystal  and the weak layer bonding are typically understood as responsible for the quasi 2D properties of these materials.  However, attesting to the important role of the interlayer coupling, it was shown that when thinned down from bulk to only a few atomic layers some of these material properties change.  For example, MoS$_2$ is in bulk an indirect-band semiconductor while one layer of the same material has a direct band gap \cite{PhysRevLett.105.136805}. Experimentally, exploring the questions of how thickness can be used to tune material properties is possible due to advances in fabrication techniques following the demonstration of micromechanical cleavage of atomically thin materials \cite{geim2013van}. In particular, the precise nature of the driving mechanisms as well as the nature of the charge density wave state in atomically thin layered materials is yet to be understood. 

\begin{figure}[htb]
\includegraphics [width=\textwidth]{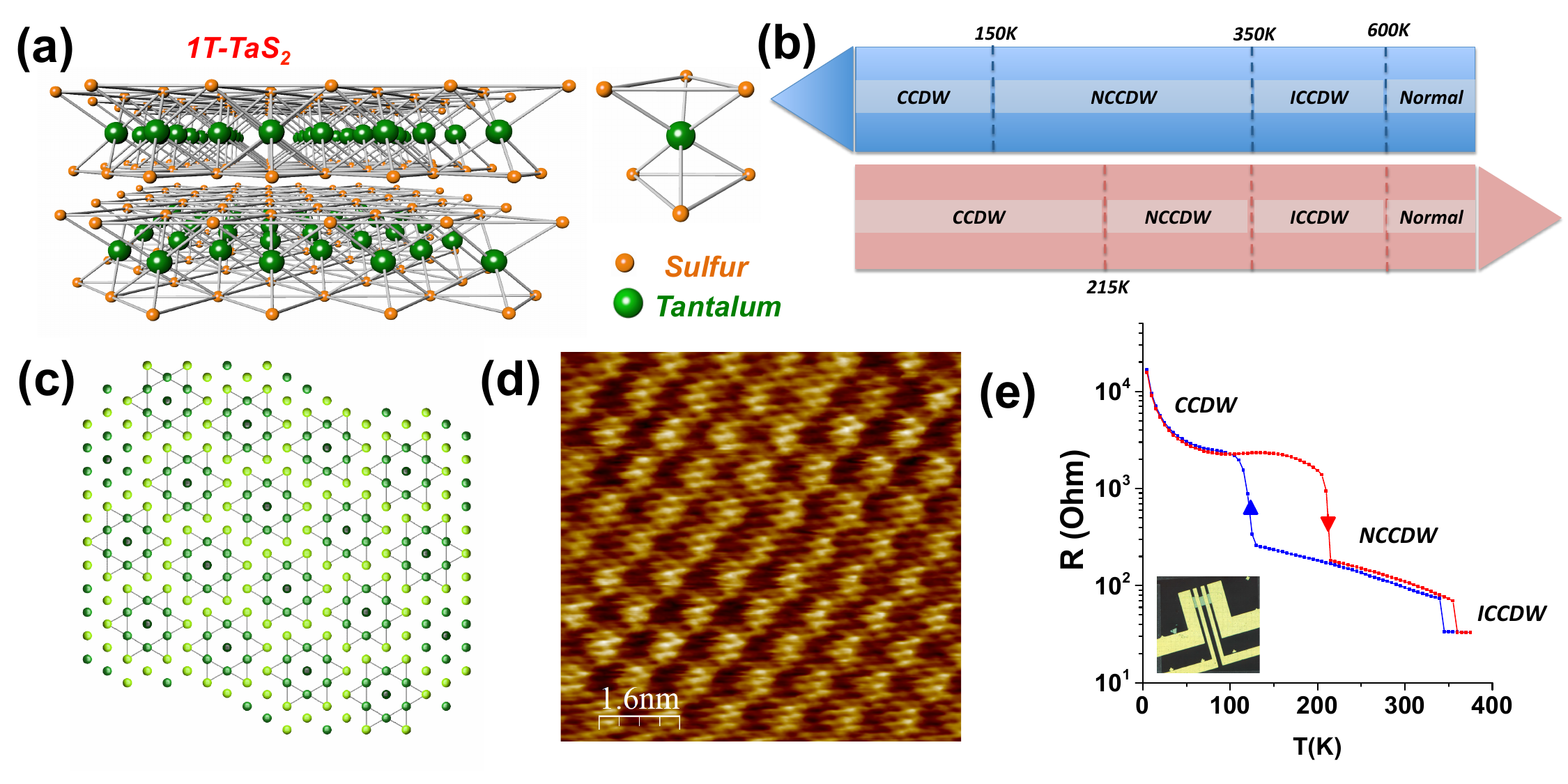}
 \caption{\label{fig1}(a) Schematic representation of the crystal structure for 1T-TaS$_2$. (b) Illustration of the different phases in 1T-TaS$_2$ as a function of temperature. (c) Ta atoms arranged in the Star of David pattern in the CCDW phase. (d) STM  topographic images of a bulk crystal of 1T-TaS$_2$  (T=59K, V$_{Bias} = 20mV$, I$_{tunneling}=2nA$) (e) Temperature dependence of the resistance of the  1T-TaS$_2$  device indicated in the inset for both the cooling (blue) and warming (red) cycles.}
\end{figure}

\begin{figure}[htb]
\includegraphics [width=125mm]{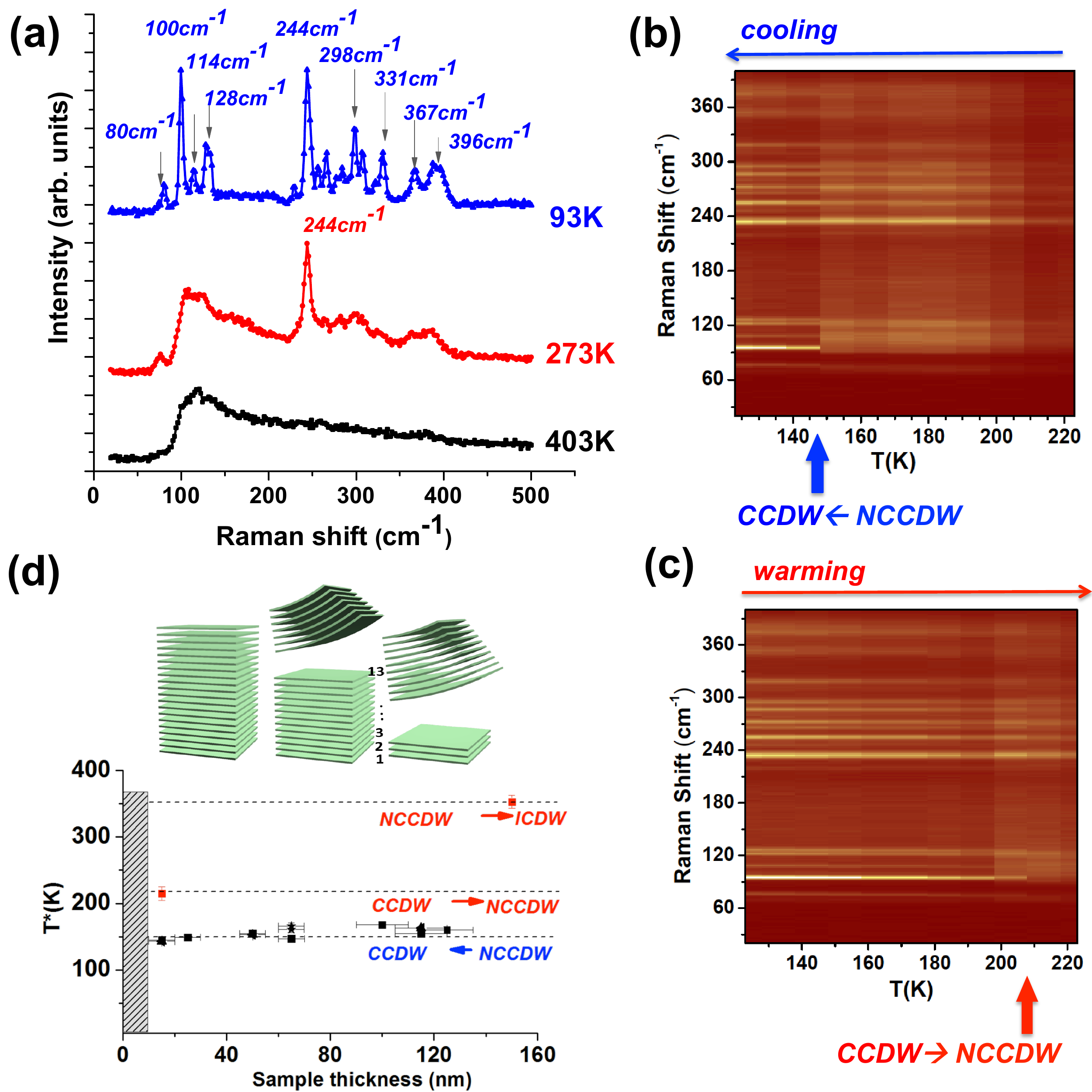}
\caption{\label{fig2} Temperature dependent Raman spectroscopy of a bulk crystal 1T-TaS$_2$. (a) Raman spectra on  a bulk sample at the indicated temperatures corresponding to different CDW phases (blue - CCDW, red - NCCDW, black - ICCDW). The red and blue curve were recorded during a cooling cycle, while the black one is on a warming cycle. (b) and (c) Raman intensity map as a function of temperature for a thick sample in a cooling and  warming cycle respectively. The sharp changes in the spectra represent the transitions indicated below the maps. (d) Top: Schematic of thinning down a crystal from bulk to a few layers. Bottom: Summary of  the transition temperatures measured for different samples as a function of sample thickness. The hatched area represent a regime where the behavior was found to be distinct from bulk. }
\end{figure}

The 1T polymorph of TaS$_2$, 1T-TaS$_2$, has one of the richest phase diagrams: it is metallic at higher temperatures, has four temperature-dependent charge density wave phases with different structures; under pressure and doping it becomes superconducting \cite{sipos2008mott} and it was suggested to show Mott insulator behavior \cite{faztos}. In this letter we use variable temperature Raman spectroscopy measurements to show that by reducing the thickness of 1T-TaS$_2$ between $\approx5nm-10nm$, the transition to a commensurate charge density wave is suppressed down to the experimentally accessed temperatures. Furthermore, the Raman spectrum suggests that for the crystals with thickness below $\approx4nm$ the nearly commensurate phase is absent.

The structure of the 1T-TaS$_2$ crystals is such that tantalum is octahedrally coordinated to the sulfur. In the c-axis, one unit cell is represented by a three-layer S-Ta-S as sketched in Figure \ref{fig1}(a). The different CDW phases that arise upon cooling and warming respectively are illustrated in Figure \ref{fig1}(b). Cooling down from  $\approx600K$ the system develops an incommensurate CDW (ICCDW) state and further cooling to approximate 350K takes it into a nearly commensurate CDW (NCCDW) arrangement. The term incommensurate refers to the fact that the CDW wavevector is not an integer of the atomic lattice wavevector. In the NCCDW, commensurate domains coexist with incommensurate ones. The transition at an even lower temperature, $\approx150K$, is  to a commensurate CDW (CCDW). In this situation, the Ta atoms participating in the associated periodic lattice distortion (PLD) are arranged in groups of 13 atoms, such a configurations being typically referred to as the Star of David (Figure \ref{fig1}(c)). When warmed up, the system shows hysteretic behavior so that it persists into the commensurate state until $\approx215K$ as illustrated in Figure \ref{fig1}(b) \cite{rossnagel2011origin}.

Flakes thicker than $20nm$ were characterized using resistivity measurements  \cite{zhang2015gate,yoshida2014controlling,tsen2015structure}.  The samples were obtained by mechanical exfoliation from a bulk crystal of TaS$_2$ onto the surface of a $Si/SiO_2$ wafer, similarly to graphene exfoliation. An optical microscope was used to identify flakes of different thicknesses and  atomic force microscope measurements were performed to determine the thickness. Typically, we obtained flakes with heights ranging from $2nm$ to over $150nm$. The devices were fabricated using a Laser Pattern Generator, photoresist as a lithographic mask and the contacts were thermally evaporated Ti/Au. An optical micrograph image of such a four-terminal device is shown as an inset to Figure \ref{fig1}(e). The resistance was recorded as a function of temperature as presented in Figure \ref{fig1}(e). The  arrows indicate both the cooling cycle (blue curve) and the heating cycle (red curve).  In the cooling cycle, the two jumps in resistance correspond to the transitions:  ICCDW-NCCDW ( $\approx$ 350$^{o}$ ) and NCCDW-CCDW ($\approx$ 150$^{o}$)  respectively. When warmed up the CDDW-NCCDW transition occurs at $\approx$ 220$^{o}$. 

\begin{figure}[htb]
\includegraphics [width=\textwidth]{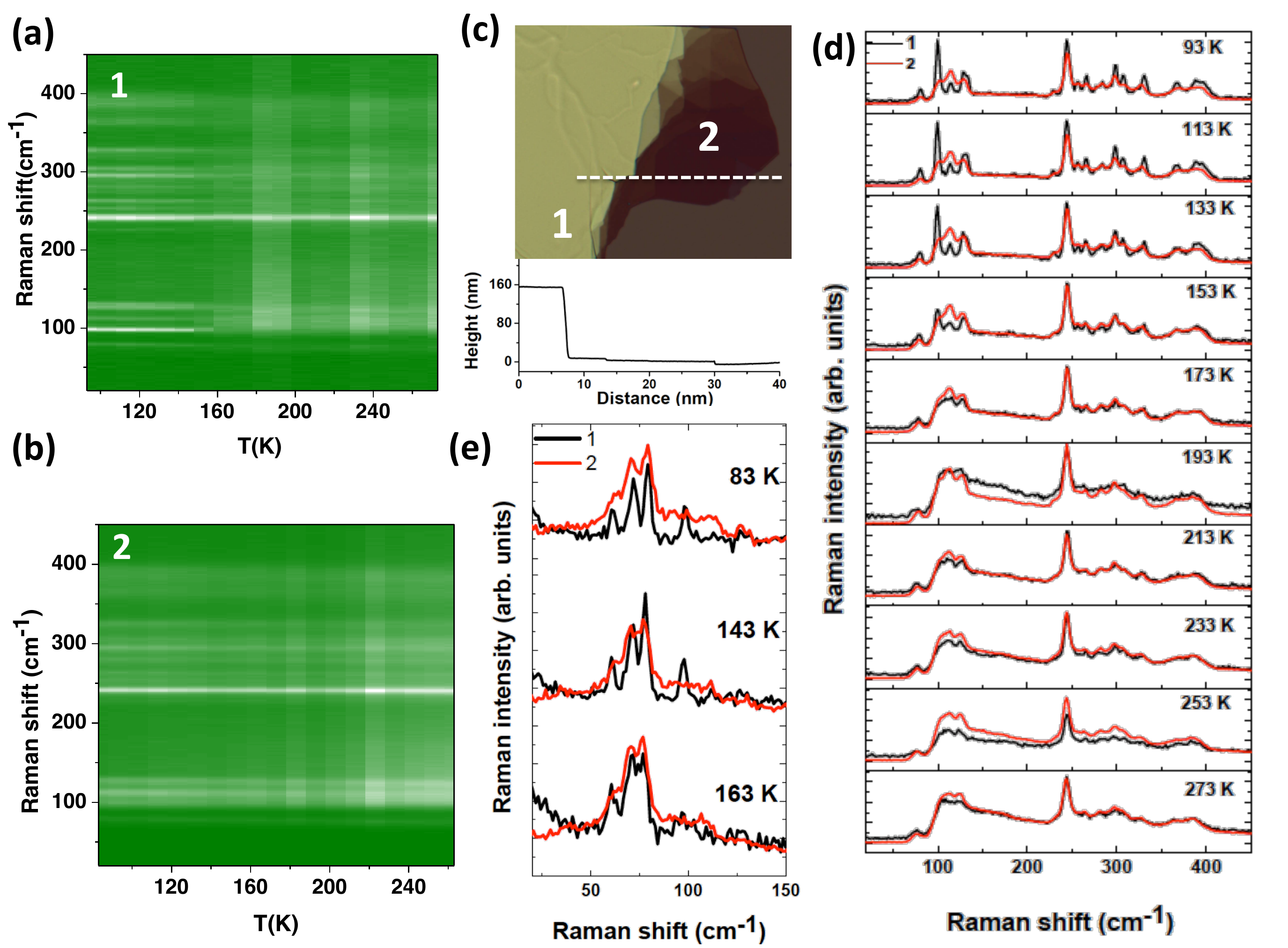}
\caption{\label{fig3} (a) Raman intensity map taken during a cooling cycle for a thick flake (h $\approx$ 160nm) indicated in (c) by number 1. (b) Raman intensity map taken during a cooling cycle for a thin flake (h $\approx$  (8 $\pm$ 2) nm)  indicated in (c) by number 2. (c) Top: Optical microscope image of the 1T-TaS$_2$ flakes where the data in (a) and (b) were taken. Bottom: Atomic force microscope height profile talken along the dashed line indicated above. (d) Comparison at the indicated temperatures between the Raman spectra taken for areas 1 and 2. (e) Raman spectra for the low wavenumbers at indicated temperatures for the areas 1 and 2. }
\end{figure}

Microscopic details of the CDW phases in 1T-TaS$_2$ can be visualized by Scanning Tunneling Microscopy (STM) \cite{burk1991charge}. Using a variable temperature ultrahigh vacuum STM the CDW on a freshly cleaved crystal was readily imaged (Figure \ref {fig1}(d)). When zooming in, both the atomic lattice as well as the CDW is visible as it is the case presented in Figure \ref{fig1}(d) which was taken at 59K.  We find the period of the CDW, in agreement with previous literature. We note that the surface becomes degraded after exposure to air for more than 30 minutes, therefore the best conditions for a tunneling experiment were achieved  by cleaving the crystal in-situ in the UHV system.

Both techniques described before complement each other in unveiling the bulk electronic and microscopic details of the CDW phases. However, they become challenging for measuring flakes with only a few atomic layers. Resistivity measurements are hampered by high contact resistance, while STM measurements on exfoliated thin flakes would also involve nanofabrication techniques that contaminate the surface. An experimental method  that can give insight into the CDW phases without the need for further processing of the thin exfoliated flakes is Raman spectroscopy. In this work we study the changes in the Raman scattering that correspond to transitions of the crystals into CDW phases, therefore probing the lattice vibrations for both the only few-layers-thick crystals as well as for the bulk. While bulk crystals were studied before \cite {Duffey1976617, Sugai1981405, Gasparov}, the experimental results presented here for atomically thin flakes have not been reported previously.

 For each selected area the Raman spectrum is recorded as a function of temperature. We use a Renishaw InVia system with the laser 514nm and typical power below 1mW .  The system is equipped with a temperature controlled stage that allowed varying the temperature of the sample from 80K to 400K. The laser spot typically focuses on a diameter of $\approx3 \mu m$. It is known that thin flakes could be damaged by heating while shining the laser light during the Raman measurement  \cite{goli2012charge}.  For the experiments here,  we checked the dependence of the spectrum on laser power and we confirmed that heating  the thin flakes can be neglected for the laser power that was used in the experiment. The same checks ensure that the temperature of the flake was not affected by shining the laser light on them.

For a bulk system we measured the Raman spectra at three different temperatures $T= 403K$, $T=233K$  and $T=93K$ respectively as shown in  Figure \ref {fig2}(a) . They have distinct features and they correspond to the ICCDW, NCCDW and CCDW phase respectively. We note that for this data set, due to the experimental set-up, the intensity of Raman spectrum below $\approx$100$ cm^{-1}$ is strongly diminished. In the NCCDW phase the most pronounced phonon modes are seen at  108 cm$^{-1}$, 124 cm$^{-1}$, 244 cm$^{-1}$, 265 cm$^{-1}$, 282 cm$^{-1}$, 300 cm$^{-1}$, 307 cm$^{-1}$, 327 cm$^{-1}$, 385 cm$^{-1}$.  At lower temperatures, for the CCDW phase, we observe that the peaks become sharper and new phonon modes appear. The appearance of new peaks can be understood as a consequence of dispersion curves folding into the zone center due to the reduced size of the Brillouin zone. In this state, because the unit cell of the CCDW phase contains 39 atoms, one expects 117 phonon modes from which $19A_g+19E_g$ \  are Raman active \cite{uchida1981infrared,Gasparov}. In our data, the Raman spectrum shows peaks at: 61 cm$^{-1}$, 71 cm$^{-1}$, 79 cm$^{-1}$,  100 cm$^{-1}$, 114 cm$^{-1}$, 128 cm$^{-1}$, 256 cm$^{-1}$, 267 cm$^{-1}$, 277 cm$^{-1}$, 284 cm$^{-1}$, 298 cm$^{-1}$, 307 cm$^{-1}$, 322 cm$^{-1}$, 331 cm$^{-1}$, 367 cm$^{-1}$, 388 cm$^{-1}$, 397 cm$^{-1}$. The most pronounced ones of the above list are also indicated by arrows in Figure \ref {fig2}(a).

To illustrate the changes in the Raman spectrum as the system goes through the different transitions, we plot Raman intensity maps as a function of temperature as presented in Figure \ref {fig2}(b) and (c). A sequence of spectra in a cooling cycle is presented in Figure \ref {fig2}(b). The change in the spectrum occurs at T $\approx 150K$ corresponding to the NCCDW-CCDW transition. In Figure \ref {fig2}(c), for the same area as in Figure \ref {fig2}(c), the temperature dependence of Raman spectrum was recorded upon warming up the system. The abrupt change corresponding to the transition CCDW-NCCDW is now present at $\approx 215K$. The hysteretic nature of this transition (Figure \ref{fig1}(e)) can, therefore,  be confirmed using the Raman spectroscopy experiment. We observe that the modes significantly affected by the transition NCCDW-CCDW are the ones around 100 $cm^{-1}$.

\begin{figure}[htb]
\includegraphics [width=\textwidth]{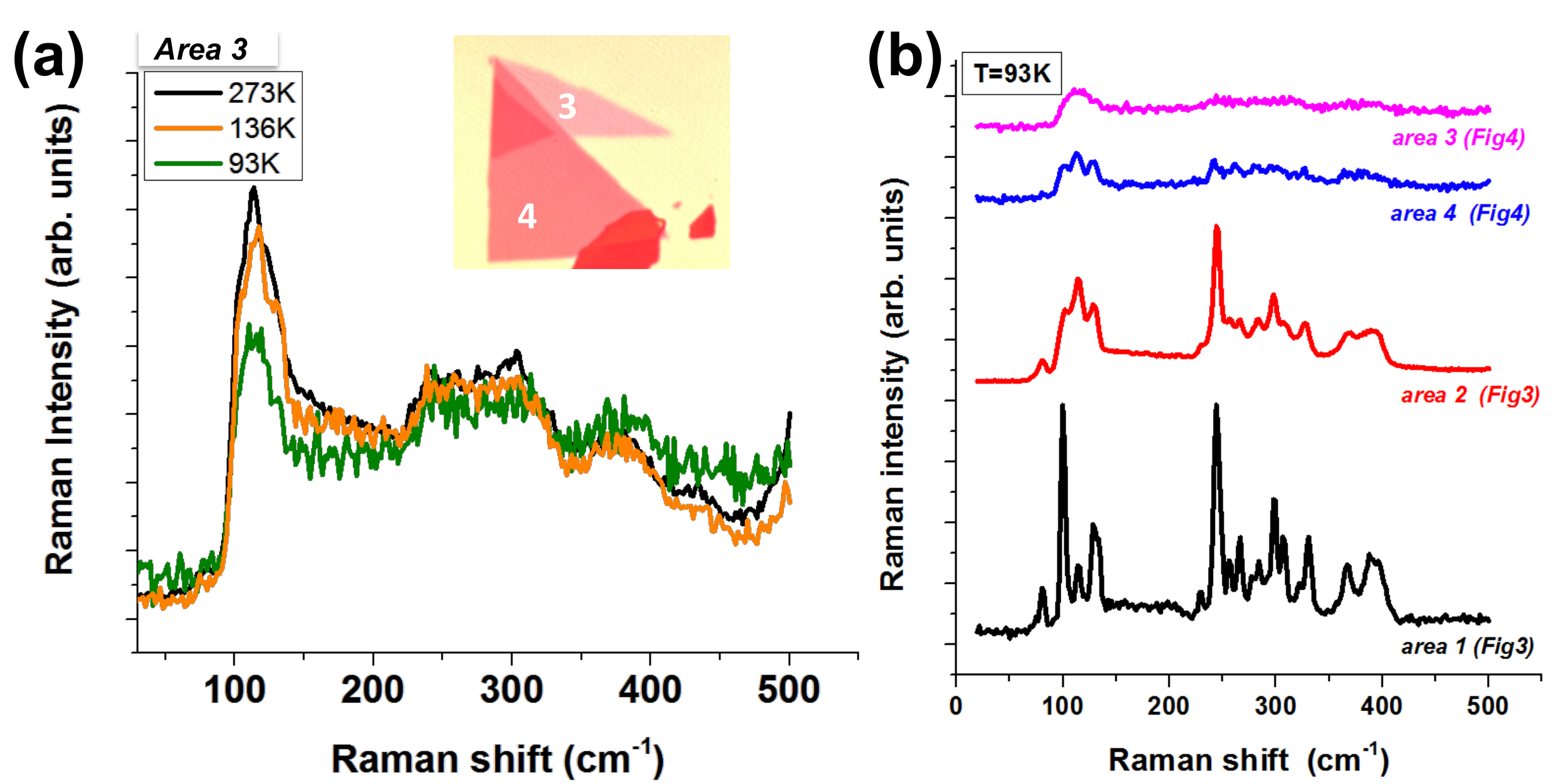}
\caption{\label{fig4} (a) Raman spectrum taken at different temperatures, at the position indicated in the optical image of the flake as area 3.   (b) Raman spectra for the indicated flakes of different thickness.  }
\end{figure}

We used such temperature dependent Raman maps as a way to explore how the different CDW phases change when thinning down the system to a few atomic layers. A summary of the experimental data is presented in Figure \ref {fig2}(d). While for flakes thicker than $\approx 10nm$  the system behaves similarly to the bulk, in the regime below $h=10nm$ the behavior of the flakes was found to be  significantly altered as explained below.

The inset of  Figure \ref {fig3}(c) shows a mechanically exfoliated flake that has a thicker part (yellow color), $h_1$ $\approx 160nm$ and a thinner part (black color) $h_2$$\approx$(8 $\pm$ 2 )nm together with a height profile obtained by atomic force microscopy.  For the two different  areas of a sample, Figures \ref {fig3}(a) and (b) present the Raman intensity map upon cooling. The thick flake of area 1 shows the abrupt change as the system transitions from NCCDW to the CCDW as explained earlier. In contrast, the thin flake of area 2, shows no transition into a CCDW, but it remains in the NCCDW at the lowest temperatures that were reached experimentally. For further comparison, in Figure \ref {fig3}(d) we plot the Raman spectra obtained on the areas 1 and 2 at the temperatures indicated as inset. Consistently, when we experimentally probe the wavenumbers below 100cm$^{-1}$ as plotted in Figure \ref {fig3}(e) we find the same behavior: while the Raman spectrum of the thick flake changed below the transition temperature, the one for the thin flake remains essentially identical to the one at higher temperatures. 

We now focus on samples that were below $ \approx 4nm$. Those are typically transparent when seen through an optical microscope as illustrated in the inset of  Figure \ref {fig4}(a). We measured the  Raman spectrum of theses atomically thin flakes both at room temperature as well as low temperature. In Figure \ref {fig4}(a) we plot the Raman spectrum taken for the flake indicated in the inset by area 3 at $T=273K, 136K, 93K$. We find that this spectrum is distinct from the one that corresponds to the NCCDW from Figure \ref {fig2}(a) at all three temperatures, indicative that that for such thin flakes the NCCDW phase is also absent. When we measure flakes of intermediate thickness such as area 4, as seen in  Figure \ref {fig4}(b), the spectrum develops a stronger peak at $\approx 244 cm^{-1}$,  suggesting that  as a function of thickness the systems undergoes novel CDW reconstructions. 

The suppression of the CCDW and NCCDW phases for the ultrathin samples is strongly connected to the periodicity of the vertical stacking corresponding to the two phases. For the NCCDW, Xray experiments have determined the presence of a 3-layer period \cite{scruby1975role, PhysRevB.56.13757}. In the case of the CCDW phase some reports propose a disordered stacking \cite{tanda1984x} while others suggest a period of 13 layers. Theoretically, various models explore what are the energetically favorable stacking arrangements in the different phases \cite{nakanishi1978domain,walker1983stacking}.
The critical thickness found by our experiments to influence the formation of the CDW phases is therefore consistent with these periods. The diminished CCDW below $\approx10nm$ and NCCDW below $\approx 4nm$ can, therefore, be understood as an effect of the crystal thickness becoming comparable to the c-axis period of the respective phase. 

In conclusion, using temperature dependent Raman spectroscopy we address the question of how the transition from bulk to few atomic layers affects the charge density wave phases in 1T-TaS$_2$. We find two thickness where the CDW phases of the crystals are significantly altered. The NCCDW is suppressed below $\approx 4nm$ and the CCDW below $\approx 10nm$ respectively.  The findings are consistent with the  previously reported c-axis periods for the two phases. These experimental results attest to the importance of the interlayer coupling to determine the properties of ultrathin van der Waals materials and open up possibilities for tuning the phase diagram of materials by varying their thickness.

\begin{acknowledgments}
Use of the Center for Nanoscale Materials was supported by the U.S. Department of Energy, Office of Science, Office of Basic Energy Sciences, under Contract No. DE-AC02- 06CH11357.
\end{acknowledgments}


\begin{thebibliography}{20}%
\makeatletter
\providecommand \@ifxundefined [1]{%
 \@ifx{#1\undefined}
}%
\providecommand \@ifnum [1]{%
 \ifnum #1\expandafter \@firstoftwo
 \else \expandafter \@secondoftwo
 \fi
}%
\providecommand \@ifx [1]{%
 \ifx #1\expandafter \@firstoftwo
 \else \expandafter \@secondoftwo
 \fi
}%
\providecommand \natexlab [1]{#1}%
\providecommand \enquote  [1]{``#1''}%
\providecommand \bibnamefont  [1]{#1}%
\providecommand \bibfnamefont [1]{#1}%
\providecommand \citenamefont [1]{#1}%
\providecommand \href@noop [0]{\@secondoftwo}%
\providecommand \href [0]{\begingroup \@sanitize@url \@href}%
\providecommand \@href[1]{\@@startlink{#1}\@@href}%
\providecommand \@@href[1]{\endgroup#1\@@endlink}%
\providecommand \@sanitize@url [0]{\catcode `\\12\catcode `\$12\catcode
  `\&12\catcode `\#12\catcode `\^12\catcode `\_12\catcode `\%12\relax}%
\providecommand \@@startlink[1]{}%
\providecommand \@@endlink[0]{}%
\providecommand \url  [0]{\begingroup\@sanitize@url \@url }%
\providecommand \@url [1]{\endgroup\@href {#1}{\urlprefix }}%
\providecommand \urlprefix  [0]{URL }%
\providecommand \Eprint [0]{\href }%
\providecommand \doibase [0]{http://dx.doi.org/}%
\providecommand \selectlanguage [0]{\@gobble}%
\providecommand \bibinfo  [0]{\@secondoftwo}%
\providecommand \bibfield  [0]{\@secondoftwo}%
\providecommand \translation [1]{[#1]}%
\providecommand \BibitemOpen [0]{}%
\providecommand \bibitemStop [0]{}%
\providecommand \bibitemNoStop [0]{.\EOS\space}%
\providecommand \EOS [0]{\spacefactor3000\relax}%
\providecommand \BibitemShut  [1]{\csname bibitem#1\endcsname}%
\let\auto@bib@innerbib\@empty
\bibitem [{\citenamefont {Wilson}\ and\ \citenamefont {Yoffe}(1969)}]{wilyof}%
  \BibitemOpen
  \bibfield  {author} {\bibinfo {author} {\bibfnamefont {J.}~\bibnamefont
  {Wilson}}\ and\ \bibinfo {author} {\bibfnamefont {A.}~\bibnamefont {Yoffe}},\
  }\href {\doibase 10.1080/00018736900101307} {\bibfield  {journal} {\bibinfo
  {journal} {Advances in Physics}\ }\textbf {\bibinfo {volume} {18}},\ \bibinfo
  {pages} {193} (\bibinfo {year} {1969})}\BibitemShut {NoStop}%
\bibitem [{\citenamefont {Mak}\ \emph {et~al.}(2010)\citenamefont {Mak},
  \citenamefont {Lee}, \citenamefont {Hone}, \citenamefont {Shan},\ and\
  \citenamefont {Heinz}}]{PhysRevLett.105.136805}%
  \BibitemOpen
  \bibfield  {author} {\bibinfo {author} {\bibfnamefont {K.~F.}\ \bibnamefont
  {Mak}}, \bibinfo {author} {\bibfnamefont {C.}~\bibnamefont {Lee}}, \bibinfo
  {author} {\bibfnamefont {J.}~\bibnamefont {Hone}}, \bibinfo {author}
  {\bibfnamefont {J.}~\bibnamefont {Shan}}, \ and\ \bibinfo {author}
  {\bibfnamefont {T.~F.}\ \bibnamefont {Heinz}},\ }\href {\doibase
  10.1103/PhysRevLett.105.136805} {\bibfield  {journal} {\bibinfo  {journal}
  {Phys. Rev. Lett.}\ }\textbf {\bibinfo {volume} {105}},\ \bibinfo {pages}
  {136805} (\bibinfo {year} {2010})}\BibitemShut {NoStop}%
\bibitem [{\citenamefont {Geim}\ and\ \citenamefont
  {Grigorieva}(2013)}]{geim2013van}%
  \BibitemOpen
  \bibfield  {author} {\bibinfo {author} {\bibfnamefont {A.}~\bibnamefont
  {Geim}}\ and\ \bibinfo {author} {\bibfnamefont {I.}~\bibnamefont
  {Grigorieva}},\ }\href@noop {} {\bibfield  {journal} {\bibinfo  {journal}
  {Nature}\ }\textbf {\bibinfo {volume} {499}},\ \bibinfo {pages} {419}
  (\bibinfo {year} {2013})}\BibitemShut {NoStop}%
\bibitem [{\citenamefont {Sipos}\ \emph {et~al.}(2008)\citenamefont {Sipos},
  \citenamefont {Kusmartseva}, \citenamefont {Akrap}, \citenamefont {Berger},
  \citenamefont {Forr{\'o}},\ and\ \citenamefont
  {Tuti{\v{s}}}}]{sipos2008mott}%
  \BibitemOpen
  \bibfield  {author} {\bibinfo {author} {\bibfnamefont {B.}~\bibnamefont
  {Sipos}}, \bibinfo {author} {\bibfnamefont {A.~F.}\ \bibnamefont
  {Kusmartseva}}, \bibinfo {author} {\bibfnamefont {A.}~\bibnamefont {Akrap}},
  \bibinfo {author} {\bibfnamefont {H.}~\bibnamefont {Berger}}, \bibinfo
  {author} {\bibfnamefont {L.}~\bibnamefont {Forr{\'o}}}, \ and\ \bibinfo
  {author} {\bibfnamefont {E.}~\bibnamefont {Tuti{\v{s}}}},\ }\href@noop {}
  {\bibfield  {journal} {\bibinfo  {journal} {Nature materials}\ }\textbf
  {\bibinfo {volume} {7}},\ \bibinfo {pages} {960} (\bibinfo {year}
  {2008})}\BibitemShut {NoStop}%
\bibitem [{\citenamefont {Fazekas}\ and\ \citenamefont
  {Tosatti}(1979)}]{faztos}%
  \BibitemOpen
  \bibfield  {author} {\bibinfo {author} {\bibfnamefont {P.}~\bibnamefont
  {Fazekas}}\ and\ \bibinfo {author} {\bibfnamefont {E.}~\bibnamefont
  {Tosatti}},\ }\href {\doibase 10.1080/13642817908245359} {\bibfield
  {journal} {\bibinfo  {journal} {Philosophical Magazine Part B}\ }\textbf
  {\bibinfo {volume} {39}},\ \bibinfo {pages} {229} (\bibinfo {year}
  {1979})}\BibitemShut {NoStop}%
\bibitem [{\citenamefont {Rossnagel}(2011)}]{rossnagel2011origin}%
  \BibitemOpen
  \bibfield  {author} {\bibinfo {author} {\bibfnamefont {K.}~\bibnamefont
  {Rossnagel}},\ }\href@noop {} {\bibfield  {journal} {\bibinfo  {journal}
  {Journal of Physics: Condensed Matter}\ }\textbf {\bibinfo {volume} {23}},\
  \bibinfo {pages} {213001} (\bibinfo {year} {2011})}\BibitemShut {NoStop}%
\bibitem [{\citenamefont {Yu}\ \emph {et~al.}(2015)\citenamefont {Yu},
  \citenamefont {Yang}, \citenamefont {Lu}, \citenamefont {Yan}, \citenamefont
  {Cho}, \citenamefont {Ma}, \citenamefont {Niu}, \citenamefont {Kim},
  \citenamefont {Son}, \citenamefont {Feng}, \citenamefont {Li}, \citenamefont
  {Cheong}, \citenamefont {Chen},\ and\ \citenamefont {Zhang}}]{zhang2015gate}%
  \BibitemOpen
  \bibfield  {author} {\bibinfo {author} {\bibfnamefont {Y.}~\bibnamefont
  {Yu}}, \bibinfo {author} {\bibfnamefont {F.}~\bibnamefont {Yang}}, \bibinfo
  {author} {\bibfnamefont {X.~F.}\ \bibnamefont {Lu}}, \bibinfo {author}
  {\bibfnamefont {Y.~J.}\ \bibnamefont {Yan}}, \bibinfo {author} {\bibfnamefont
  {Y.-H.}\ \bibnamefont {Cho}}, \bibinfo {author} {\bibfnamefont
  {L.}~\bibnamefont {Ma}}, \bibinfo {author} {\bibfnamefont {X.}~\bibnamefont
  {Niu}}, \bibinfo {author} {\bibfnamefont {S.}~\bibnamefont {Kim}}, \bibinfo
  {author} {\bibfnamefont {Y.-W.}\ \bibnamefont {Son}}, \bibinfo {author}
  {\bibfnamefont {D.}~\bibnamefont {Feng}}, \bibinfo {author} {\bibfnamefont
  {S.}~\bibnamefont {Li}}, \bibinfo {author} {\bibfnamefont {S.-W.}\
  \bibnamefont {Cheong}}, \bibinfo {author} {\bibfnamefont {X.~H.}\
  \bibnamefont {Chen}}, \ and\ \bibinfo {author} {\bibfnamefont
  {Y.}~\bibnamefont {Zhang}},\ }\href@noop {} {\bibfield  {journal} {\bibinfo
  {journal} {Nature Nanotechnology}\ }\textbf {\bibinfo {volume} {10}},\
  \bibinfo {pages} {270} (\bibinfo {year} {2015})}\BibitemShut {NoStop}%
\bibitem [{\citenamefont {Yoshida}\ \emph {et~al.}(2014)\citenamefont
  {Yoshida}, \citenamefont {Zhang}, \citenamefont {Ye}, \citenamefont {Suzuki},
  \citenamefont {Imai}, \citenamefont {Kimura}, \citenamefont {Fujiwara},\ and\
  \citenamefont {Iwasa}}]{yoshida2014controlling}%
  \BibitemOpen
  \bibfield  {author} {\bibinfo {author} {\bibfnamefont {M.}~\bibnamefont
  {Yoshida}}, \bibinfo {author} {\bibfnamefont {Y.}~\bibnamefont {Zhang}},
  \bibinfo {author} {\bibfnamefont {J.}~\bibnamefont {Ye}}, \bibinfo {author}
  {\bibfnamefont {R.}~\bibnamefont {Suzuki}}, \bibinfo {author} {\bibfnamefont
  {Y.}~\bibnamefont {Imai}}, \bibinfo {author} {\bibfnamefont {S.}~\bibnamefont
  {Kimura}}, \bibinfo {author} {\bibfnamefont {A.}~\bibnamefont {Fujiwara}}, \
  and\ \bibinfo {author} {\bibfnamefont {Y.}~\bibnamefont {Iwasa}},\
  }\href@noop {} {\bibfield  {journal} {\bibinfo  {journal} {Scientific
  reports}\ }\textbf {\bibinfo {volume} {4}} (\bibinfo {year}
  {2014})}\BibitemShut {NoStop}%
\bibitem [{\citenamefont {Tsen}\ \emph {et~al.}(2015)\citenamefont {Tsen},
  \citenamefont {Hovden}, \citenamefont {Wang}, \citenamefont {Kim},
  \citenamefont {Okamoto}, \citenamefont {Spoth}, \citenamefont {Liu},
  \citenamefont {Lu}, \citenamefont {Sun}, \citenamefont {Hone} \emph
  {et~al.}}]{tsen2015structure}%
  \BibitemOpen
  \bibfield  {author} {\bibinfo {author} {\bibfnamefont {A.}~\bibnamefont
  {Tsen}}, \bibinfo {author} {\bibfnamefont {R.}~\bibnamefont {Hovden}},
  \bibinfo {author} {\bibfnamefont {D.}~\bibnamefont {Wang}}, \bibinfo {author}
  {\bibfnamefont {Y.}~\bibnamefont {Kim}}, \bibinfo {author} {\bibfnamefont
  {J.}~\bibnamefont {Okamoto}}, \bibinfo {author} {\bibfnamefont
  {K.}~\bibnamefont {Spoth}}, \bibinfo {author} {\bibfnamefont
  {Y.}~\bibnamefont {Liu}}, \bibinfo {author} {\bibfnamefont {W.}~\bibnamefont
  {Lu}}, \bibinfo {author} {\bibfnamefont {Y.}~\bibnamefont {Sun}}, \bibinfo
  {author} {\bibfnamefont {J.}~\bibnamefont {Hone}},  \emph {et~al.},\
  }\href@noop {} {\bibfield  {journal} {\bibinfo  {journal} {arXiv preprint
  arXiv:1505.03769}\ } (\bibinfo {year} {2015})}\BibitemShut {NoStop}%
\bibitem [{\citenamefont {Burk}\ \emph {et~al.}(1991)\citenamefont {Burk},
  \citenamefont {Thomson}, \citenamefont {Zettl},\ and\ \citenamefont
  {Clarke}}]{burk1991charge}%
  \BibitemOpen
  \bibfield  {author} {\bibinfo {author} {\bibfnamefont {B.}~\bibnamefont
  {Burk}}, \bibinfo {author} {\bibfnamefont {R.}~\bibnamefont {Thomson}},
  \bibinfo {author} {\bibfnamefont {A.}~\bibnamefont {Zettl}}, \ and\ \bibinfo
  {author} {\bibfnamefont {J.}~\bibnamefont {Clarke}},\ }\href@noop {}
  {\bibfield  {journal} {\bibinfo  {journal} {Physical review letters}\
  }\textbf {\bibinfo {volume} {66}},\ \bibinfo {pages} {3040} (\bibinfo {year}
  {1991})}\BibitemShut {NoStop}%
\bibitem [{\citenamefont {Duffey}\ \emph {et~al.}(1976)\citenamefont {Duffey},
  \citenamefont {Kirby},\ and\ \citenamefont {Coleman}}]{Duffey1976617}%
  \BibitemOpen
  \bibfield  {author} {\bibinfo {author} {\bibfnamefont {J.}~\bibnamefont
  {Duffey}}, \bibinfo {author} {\bibfnamefont {R.}~\bibnamefont {Kirby}}, \
  and\ \bibinfo {author} {\bibfnamefont {R.}~\bibnamefont {Coleman}},\ }\href
  {\doibase http://dx.doi.org/10.1016/0038-1098(76)91073-5} {\bibfield
  {journal} {\bibinfo  {journal} {Solid State Communications}\ }\textbf
  {\bibinfo {volume} {20}},\ \bibinfo {pages} {617 } (\bibinfo {year}
  {1976})}\BibitemShut {NoStop}%
\bibitem [{\citenamefont {Sugai}\ \emph {et~al.}(1981)\citenamefont {Sugai},
  \citenamefont {Murase}, \citenamefont {Uchida},\ and\ \citenamefont
  {Tanaka}}]{Sugai1981405}%
  \BibitemOpen
  \bibfield  {author} {\bibinfo {author} {\bibfnamefont {S.}~\bibnamefont
  {Sugai}}, \bibinfo {author} {\bibfnamefont {K.}~\bibnamefont {Murase}},
  \bibinfo {author} {\bibfnamefont {S.}~\bibnamefont {Uchida}}, \ and\ \bibinfo
  {author} {\bibfnamefont {S.}~\bibnamefont {Tanaka}},\ }\href {\doibase
  http://dx.doi.org/10.1016/0378-4363(81)90284-9} {\bibfield  {journal}
  {\bibinfo  {journal} {Physica B+C}\ }\textbf {\bibinfo {volume} {105}},\
  \bibinfo {pages} {405 } (\bibinfo {year} {1981})}\BibitemShut {NoStop}%
\bibitem [{\citenamefont {Gasparov}\ \emph {et~al.}(2002)\citenamefont
  {Gasparov}, \citenamefont {Brown}, \citenamefont {Wint}, \citenamefont
  {Tanner}, \citenamefont {Berger}, \citenamefont {Margaritondo}, \citenamefont
  {Ga\'al},\ and\ \citenamefont {Forr\'o}}]{Gasparov}%
  \BibitemOpen
  \bibfield  {author} {\bibinfo {author} {\bibfnamefont {L.~V.}\ \bibnamefont
  {Gasparov}}, \bibinfo {author} {\bibfnamefont {K.~G.}\ \bibnamefont {Brown}},
  \bibinfo {author} {\bibfnamefont {A.~C.}\ \bibnamefont {Wint}}, \bibinfo
  {author} {\bibfnamefont {D.~B.}\ \bibnamefont {Tanner}}, \bibinfo {author}
  {\bibfnamefont {H.}~\bibnamefont {Berger}}, \bibinfo {author} {\bibfnamefont
  {G.}~\bibnamefont {Margaritondo}}, \bibinfo {author} {\bibfnamefont
  {R.}~\bibnamefont {Ga\'al}}, \ and\ \bibinfo {author} {\bibfnamefont
  {L.}~\bibnamefont {Forr\'o}},\ }\href {\doibase 10.1103/PhysRevB.66.094301}
  {\bibfield  {journal} {\bibinfo  {journal} {Phys. Rev. B}\ }\textbf {\bibinfo
  {volume} {66}},\ \bibinfo {pages} {094301} (\bibinfo {year}
  {2002})}\BibitemShut {NoStop}%
\bibitem [{\citenamefont {Goli}\ \emph {et~al.}(2012)\citenamefont {Goli},
  \citenamefont {Khan}, \citenamefont {Wickramaratne}, \citenamefont {Lake},\
  and\ \citenamefont {Balandin}}]{goli2012charge}%
  \BibitemOpen
  \bibfield  {author} {\bibinfo {author} {\bibfnamefont {P.}~\bibnamefont
  {Goli}}, \bibinfo {author} {\bibfnamefont {J.}~\bibnamefont {Khan}}, \bibinfo
  {author} {\bibfnamefont {D.}~\bibnamefont {Wickramaratne}}, \bibinfo {author}
  {\bibfnamefont {R.~K.}\ \bibnamefont {Lake}}, \ and\ \bibinfo {author}
  {\bibfnamefont {A.~A.}\ \bibnamefont {Balandin}},\ }\href@noop {} {\bibfield
  {journal} {\bibinfo  {journal} {Nano letters}\ }\textbf {\bibinfo {volume}
  {12}},\ \bibinfo {pages} {5941} (\bibinfo {year} {2012})}\BibitemShut
  {NoStop}%
\bibitem [{\citenamefont {Uchida}\ and\ \citenamefont
  {Sugai}(1981)}]{uchida1981infrared}%
  \BibitemOpen
  \bibfield  {author} {\bibinfo {author} {\bibfnamefont {S.}~\bibnamefont
  {Uchida}}\ and\ \bibinfo {author} {\bibfnamefont {S.}~\bibnamefont {Sugai}},\
  }\href@noop {} {\bibfield  {journal} {\bibinfo  {journal} {Physica B+ C}\
  }\textbf {\bibinfo {volume} {105}},\ \bibinfo {pages} {393} (\bibinfo {year}
  {1981})}\BibitemShut {NoStop}%
\bibitem [{\citenamefont {Scruby}\ \emph {et~al.}(1975)\citenamefont {Scruby},
  \citenamefont {Williams},\ and\ \citenamefont {Parry}}]{scruby1975role}%
  \BibitemOpen
  \bibfield  {author} {\bibinfo {author} {\bibfnamefont {C.}~\bibnamefont
  {Scruby}}, \bibinfo {author} {\bibfnamefont {P.}~\bibnamefont {Williams}}, \
  and\ \bibinfo {author} {\bibfnamefont {G.}~\bibnamefont {Parry}},\
  }\href@noop {} {\bibfield  {journal} {\bibinfo  {journal} {Philosophical
  Magazine}\ }\textbf {\bibinfo {volume} {31}},\ \bibinfo {pages} {255}
  (\bibinfo {year} {1975})}\BibitemShut {NoStop}%
\bibitem [{\citenamefont {Spijkerman}\ \emph {et~al.}(1997)\citenamefont
  {Spijkerman}, \citenamefont {de~Boer}, \citenamefont {Meetsma}, \citenamefont
  {Wiegers},\ and\ \citenamefont {van Smaalen}}]{PhysRevB.56.13757}%
  \BibitemOpen
  \bibfield  {author} {\bibinfo {author} {\bibfnamefont {A.}~\bibnamefont
  {Spijkerman}}, \bibinfo {author} {\bibfnamefont {J.~L.}\ \bibnamefont
  {de~Boer}}, \bibinfo {author} {\bibfnamefont {A.}~\bibnamefont {Meetsma}},
  \bibinfo {author} {\bibfnamefont {G.~A.}\ \bibnamefont {Wiegers}}, \ and\
  \bibinfo {author} {\bibfnamefont {S.}~\bibnamefont {van Smaalen}},\ }\href
  {\doibase 10.1103/PhysRevB.56.13757} {\bibfield  {journal} {\bibinfo
  {journal} {Phys. Rev. B}\ }\textbf {\bibinfo {volume} {56}},\ \bibinfo
  {pages} {13757} (\bibinfo {year} {1997})}\BibitemShut {NoStop}%
\bibitem [{\citenamefont {Tanda}\ \emph {et~al.}(1984)\citenamefont {Tanda},
  \citenamefont {Sambongi}, \citenamefont {Tani},\ and\ \citenamefont
  {Tanaka}}]{tanda1984x}%
  \BibitemOpen
  \bibfield  {author} {\bibinfo {author} {\bibfnamefont {S.}~\bibnamefont
  {Tanda}}, \bibinfo {author} {\bibfnamefont {T.}~\bibnamefont {Sambongi}},
  \bibinfo {author} {\bibfnamefont {T.}~\bibnamefont {Tani}}, \ and\ \bibinfo
  {author} {\bibfnamefont {S.}~\bibnamefont {Tanaka}},\ }\href@noop {}
  {\bibfield  {journal} {\bibinfo  {journal} {Journal of the Physical Society
  of Japan}\ }\textbf {\bibinfo {volume} {53}},\ \bibinfo {pages} {476}
  (\bibinfo {year} {1984})}\BibitemShut {NoStop}%
\bibitem [{\citenamefont {Nakanishi}\ and\ \citenamefont
  {Shiba}(1978)}]{nakanishi1978domain}%
  \BibitemOpen
  \bibfield  {author} {\bibinfo {author} {\bibfnamefont {K.}~\bibnamefont
  {Nakanishi}}\ and\ \bibinfo {author} {\bibfnamefont {H.}~\bibnamefont
  {Shiba}},\ }\href@noop {} {\bibfield  {journal} {\bibinfo  {journal} {Journal
  of the Physical Society of Japan}\ }\textbf {\bibinfo {volume} {45}},\
  \bibinfo {pages} {1147} (\bibinfo {year} {1978})}\BibitemShut {NoStop}%
\bibitem [{\citenamefont {Walker}\ and\ \citenamefont
  {Withers}(1983)}]{walker1983stacking}%
  \BibitemOpen
  \bibfield  {author} {\bibinfo {author} {\bibfnamefont {M.}~\bibnamefont
  {Walker}}\ and\ \bibinfo {author} {\bibfnamefont {R.}~\bibnamefont
  {Withers}},\ }\href@noop {} {\bibfield  {journal} {\bibinfo  {journal}
  {Physical Review B}\ }\textbf {\bibinfo {volume} {28}},\ \bibinfo {pages}
  {2766} (\bibinfo {year} {1983})}\BibitemShut {NoStop}%
\end{thebibliography}
\end{document}